\documentclass[letterpaper,twocolumn,10pt]{article}

\usepackage{amsmath}
\usepackage{usenix,epsfig,endnotes}
\usepackage[utf8]{inputenc}
\usepackage{graphicx, subfigure}
\usepackage[hyphens]{url}

\title{Reflection Scan: an Off-Path Attack on TCP}

\author{
{\rm Jan Wróbel}\\
{\small \rm \url{wrr@mixedbit.org}}
}
\begin{document}
\date{}
\maketitle

\begin{abstract}
The paper demonstrates how traffic load of a shared packet queue can
be exploited as a side channel through which protected information
leaks to an off-path attacker. The attacker sends to a victim a
sequence of identical spoofed segments. The victim responds to each
segment in the sequence (the sequence is reflected by the victim) if
the segments satisfy a certain condition tested by the attacker. The
responses do not reach the attacker directly, but induce extra load on
a routing queue shared between the victim and the attacker. Increased
processing time of packets traversing the queue reveal that the tested
condition was true. The paper concentrates on the TCP, but the
approach is generic and can be effective against other protocols that
allow to construct requests which are conditionally answered by the
victim. A proof of concept was created to assess applicability of the
method in real-life scenarios.
\end{abstract}

\section{Introduction}

The TCP protocol without an additional encryption and authentication
layer is inherently vulnerable to man-in-the-middle attacks. An
attacker that has a way to intercept network traffic between TCP end
points, can easily read and alter the communication. Off-path attacks,
in which the attacker can not intercept network traffic, are much
harder to execute. Along the years several weaknesses in the protocol
or particular implementations that made off-path attacks easier were
disclosed. Protocol specification was improved and many vendors fixed
implementations to close discovered holes. A TCP connection between
hosts that implement the newest recommendations (\cite{blind_robust},
\cite{port_randomization}, \cite{tcp_sec_assessment}) is believed to
be reasonably well protected against off-path attacks.

A TCP session is protected by three secret numbers: a 16-bit ephemeral
port and two 32-bit sequence numbers, one for each side of the
connection. Other fields, such as IP addresses of end points and a
server port, are easy to determine in many scenarios. Each TCP segment
exchanged within an established connection carries all three secret
values. For a segment to be accepted, it must contain a correct
ephemeral port number, its sequence number must be within receiver's
window and a sequence number the segment is acknowledging (acknowledge
number) must be acceptable. According to the recent recommendations,
an ephemeral port should be randomly picked from a 1025-65535 range
and an acknowledge number should be accepted only if it is equal to
the next octet to be sent or lower by at most 'largest sender window
seen'. If an end point follows these recommendations, the attacker needs
\[
\frac{
(2^{16}-1025) \times 2^{32} \times  2^{32}}
{\text{window size A} \times \text{window size B}}
\]
attempts to generate an acceptable segment. Assuming both windows have
65kB, about $2^{48}$ attempts are needed. If the end point follows
strict RST validation rules, which require RST segment to have a
sequence number equal to the next expected sequence number, the
attacker needs $(2^{16}-1025)\times2^{32}$ attempts to blindly reset
the connection, which is also about $2^{48}$. The number is large
enough to make blind attacks impractical in most scenarios. The
attacker would need to push segments for 500 hours at 100Gb/s rate to
have one segment accepted. Even if a segment is accepted, the
probability that it lines up with a start of a window is only
$\frac{1}{\text{window size}}$. Thus, a successful blind attack can
corrupt or reset the session, but it has low chances of inserting a
meaningful payload in a correct place.

While the risk of accepting spoofed TCP segments as valid is
recognized and well studied, the recommendations and implementations
overlook the risk of responding to rejected segments.  A TCP layer can
either silently drop a rejected segment or respond to it (with an ACK
or a RST). The action to perform differs between different
implementations of the protocol. It was originally specified in the
'Event Processing' section of RFC 793~\cite{rfc}, but new systems,
especially firewalls, do not fully follow the RFC, but implement
stricter filtering rules, as for example described
in~\cite{ipfilter}. These new rules are carefully specified to
preserve interoperability between different implementations.

If for a particular TCP implementation conditional response to a
rejected segment depends on one of secret values set in the segment,
and if an attacker can discover that a system responded to a spoofed
segment, the TCP session can be compromised. The attacker can
determine if a tested secret value satisfied certain condition (an
ephemeral port was correct, a sequence number was in window, an
acknowledge number was acceptable). The secrets can be revealed in
separate steps, each of the steps requires relatively small resources.

\begin{figure}[h!]
\begin{center}
\includegraphics[width=0.9\columnwidth,clip]{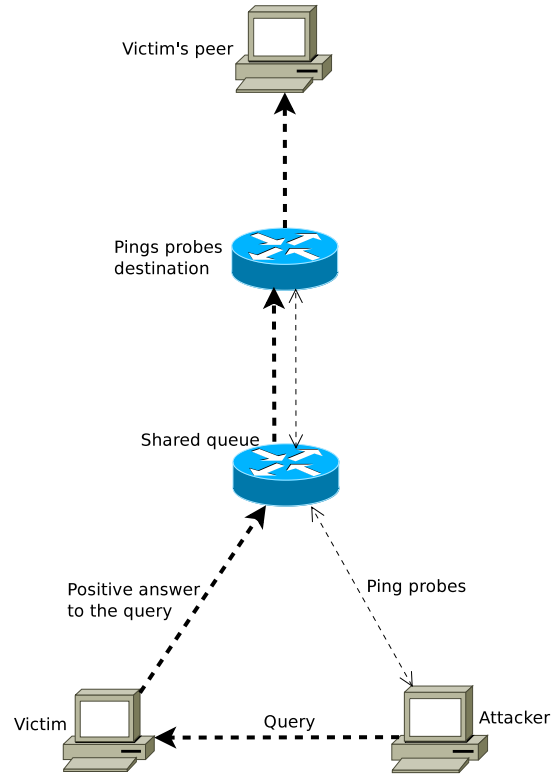}
\end{center}
\caption{High level attack scheme. The attacker sends a query to the
  victim in a form of a sequence of spoofed segments. If the answer to
  the query is positive, the victim responds with a sequence of
  segments addressed to its peer. At the same time, the attacker sends
  ping probes that share an outbound queue with segments from the
  victim. Increased round trip time reveals the positive answer to the
  query.}
\label{reflection_scan}
\end{figure}

Congestion of a queue shared between the off-path attacker and the
targeted TCP stream is a side channel through which the attacker
can determine if the TCP layer responded to spoofed
segments. Detecting negligible load caused by a single response would
be hard in practice, but the attacker can send a sequence of segments
of any desired length. If each segment from the sequence is answered,
the answers can cause a substantial traffic spike or even queue
overflow. Figure \ref{reflection_scan} illustrates the technique.

\section{Related work}

A high correlation between traffic patterns of users sharing a routing
resource was demonstrated in~\cite{side_channel}. The authors monitored ping
round trip time to a router that connected a user to the Internet and
compared the measurements with traffic patterns generated by the
user's online activities. In this technique the eavesdropper was
passive and did not send any packets to trigger traffic spikes and
gain additional information.

The attack described in this paper shares a lot of similarities with
well known off-path techniques that exploit weak implementations of IP
ID generation mechanism. Some legacy systems increase the ID field of
subsequent IP packets that leave a machine by one. This provides a
side channel to determine if a host sent a packet in response to
incoming traffic. The channel can be used to perform stealth port
scans~\cite{stealth_scan} or to execute off-path attacks against
established TCP connections~\cite{ip_id_tcp_attack}. Contrary to the
technique described in this paper, the exploitation of IP ID channel
requires an attacker to establish a legitimate, bidirectional
communication channel to a vulnerable host. Today firewalls commonly
disallow creation of such channels to client machines. This document
concentrates on compromising TCP session, but the technique can also
be used to perform a stealth port scan analogous to the one described
in~\cite{stealth_scan}.

The authors of~\cite{congestion} showed that TCP congestion control
mechanism can be exploited by a malicious receiver to improve
performance of his/her connection at the cost of others.
Applicability of such technique for a Denial of Service attack was
studied in~\cite{congestion_dos}. In simulated environment the authors
were able to significantly decrease the bandwidth of participating TCP
connections. Security Assessment of the TCP~\cite{tcp_sec_assessment}
explains that such attacks can be executed blindly by an off-path
attacker. Congestion control mechanism is driven by ACK segments and
TCP layer can be easily tricked to generate ACKs by spoofed segments
with incorrect sequence numbers.

\section{Requirements and applicability}

As in case of most off-path attacks, the attacker must be able
to send spoofed IP packets to one end of the targeted connection. It
is also assumed that IP addresses of both ends and a port number of a
server are known to the attacker. Throughout the paper, the end
point to which spoofed segments are addressed is called 'the victim', the
second end point is called 'the victim's peer'.

In addition to these usual requirements, the attacker must be able to
send legitimate traffic probes through (or to) one of the machines (a
router or an end point) on the path of the targeted TCP
traffic. Ideally, the machine should be a bottleneck for the TCP
connection. As described in~\cite{side_channel}, a good candidate is
an edge router connecting the victim to the Internet. The probes can
be ICMP pings, but also segments exchanged within a legitimate TCP
connection, anything that would allow to detect changes in traffic
load of the bottleneck.

There are various factors that influence applicability of the attack:
\begin{itemize}
\item Available bandwidth and time. The bigger the bandwidth between
  the attacker and the victim the better. The smaller the
  bandwidth between the victim and its peer the better.
\item Bottleneck's natural traffic patterns. The attack is harder if
the traffic traversing the bottleneck is large or has variable
characteristic.
\item Network topology. The attack is easier if spoofed segments from
  the attacker to the victim do not traverse the bottleneck, and thus
  do not disturb traffic probes sent by the attacker.
\item Bottleneck's queuing policy. Good isolation of traffic coming from
 different users can impede the attack.
\item Traffic measuring and analyzing technique. Advanced
  techniques can increase the attack feasibility in adverse scenarios.
\end{itemize}

It is beyond the scope of this paper to determine the practical limits
of the technique. The results of performed experiments can provide a
reference point for analyzing applicability of the attack in different
scenarios. The proof of concept can be used a starting point
for further experiments. The attack requires much fewer resources
than truly blind off-path attack, but the requirements are still
significant enough to make it impractical in many real-life scenarios.

\section{Experimental setup}

The experiments were performed in favorable for the attacker, but not
improbable conditions. The attacker was sharing an edge router with
the victim. The router had 2500kb/s downlink and 320kb/s uplink
connection to the Internet. The attacker was connected to the victim
with 100Mb/s link, but did not have direct access to the
victim's traffic. Three different scenarios were considered:
\begin{itemize}
  \item Idle TCP connection with negligible natural traffic traversing
    the bottleneck. This scenario was the easiest one, induced
    responses constituted substantial part of bottleneck's traffic.
  \item The victim downloading data at full speed (saturated downlink).
  \item The victim uploading data at full speed (saturated uplink).
\end{itemize}
The attacker sent ping requests to a router one hop beyond the edge
router. This ensured ping packets and segments sent by the victim in
response to spoofed traffic shared an outgoing queue of the edge
router.  When the link to the outside world was idle, the ping Round
Trip Time was about 20ms, when the link was saturated, the RTT
increased to about 700ms.

Two systems were analyzed. Windows XP SP3 with
firewall enabled and Linux 3.0.0. Linux had Netfilter firewall
enabled with following commands:
\begin{verbatim}
  iptables -A INPUT -m state \
    --state ESTABLISHED -j ACCEPT;
  iptables -A INPUT -j DROP;
\end{verbatim}
This is a common configuration for a client machine. All incoming
traffic that it not directed to connections initiated by the protected
machine is dropped.

The two tested systems implement different rules for processing TCP
segments. To determine how a host protected by a firewall responds to
an incoming segment two steps need to be analyzed: is firewall going
to drop the segment and if not, how TCP layer is going to handle the
segment? The differences between the two tested systems come from the
first step - Netfilter imposes stricter filtering rules
(\cite{ipfilter}) than Windows XP firewall. The second step for both
systems is the same (in respect to processing rules exploited by the
attack) and closely follows RFC 793. Processing rules that are
important from the attack perspective are briefly explained in
following sections.

The proof of concept that was used to obtain experimental results can be
found at~\cite{poc_code}. The paper does not discuss low level details
of the implementation, an interested reader is encouraged to study a
documentation accompanying the code.

It is important to note that no bugs in TCP implementations of
targeted systems were exploited.

\section{Attack details}
\label{details}
Assuming a shared router implements FIFO queuing policy, delay
introduced by a series of $N$ packets of equal size is:
\[
\frac{N * \text{packet size}}{\text{bandwith}}
\]
The victim is tricked to generate ACK segments, which have about 80
bytes (assuming about 40B for layer two header, 20B for IP and about
20B for TCP headers). Applying the formula to the experimental setup,
a theoretical delay introduced by 30 ACK segments should be
$\frac{30*80*8}{320000}\text{s}=0.06\text{s}$. It is three times more
than the ping RTT for the idle link (20ms), and should be easily
detectable in the easiest experimental scenario. 1000 ACKs should
introduce a delay of
$\frac{1000*80*8}{320000}\text{s}=2.0\text{s}$. This is about
three times more than the ping RTT for the saturated link (700ms), and
should be easily detectable in the download and upload experiments.

\subsection{Ephemeral port number}
\label{ephemeral_port}
'Event Processing' section of RFC 793 requires an ACK segment to be
sent in response to any segment that belongs to an established
connection (has correct IP addresses and ports) but is outside of a
window (has an incorrect sequence number). If a host adheres to this
specification, and is protected by a firewall that silently drops
segments not belonging to any connection (a common case), the attacker
can use segments with an incorrect sequence number to determine a
client port number.

Windows and Linux TCP stacks follow the RFC and respond with ACK to
any segment with an incorrect sequence number. Linux Netfilter firewall
uses stricter validation rules to drop segments that are not part of a
connection:
\begin{itemize}
\item Segments without ACK flag are dropped.
\item Acknowledge number is validated. It is accepted only if it is
  equal to the next octet to be sent or lower by at most $\max(66000,
  \text{largest sender window seen})$.
\end{itemize}

Acknowledge number validation makes it much harder to use segments with
an incorrect sequence number to search for a client port. But there is a hole:
\begin{itemize}
\item Segments that have both SYN and ACK flag set are always
  accepted and passed to the TCP layer.
\end{itemize}

TCP layer responds with ACK to such segments if their sequence number
is outside of a window. This allows to discover an ephemeral port of a
Netfilter protected host. The only drawback is that if a sequence
number of SYN-ACK segment accidentally happens to be in-window,
Linux responds with RST and the connection is closed. The probability
of this is low: $\frac{\text{window size}}{2^{32}}$.

\begin{figure}[h!]
\begin{center}
\subfigure[
  Connection idle, 5 pings/port, 30 spoofed segments/port]{
  \includegraphics[width=0.99\columnwidth,clip]{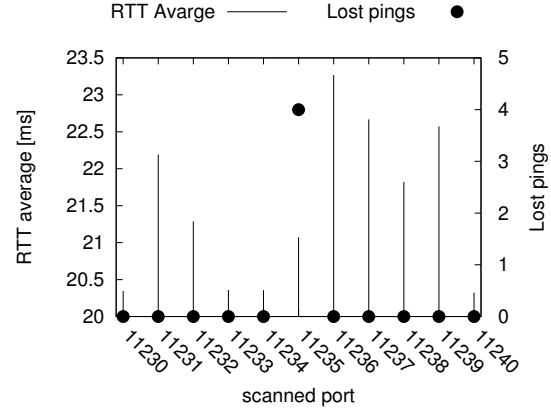}
}
\subfigure[
  Connection downloading data, 10 pings/port, 1000 spoofed segments/port]{
  \includegraphics[width=0.99\columnwidth,clip]{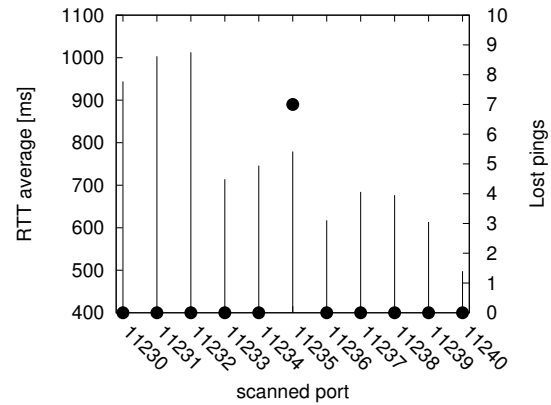}
}
\end{center}
\caption{The change in pings' loss rate reveal an ephemeral port in
  use (11235). Ping is considered lost if the response did not arrive
  within two RTTs of previous pings related to the same port.}
\label{port_scan}
\end{figure}

Figure~\ref{port_scan} shows how ping RTT increases when a sequence of
spoofed segments is directed at the correct ephemeral port. A spike in
RTT occurs reliably, but usually it is not the only detected
spike. Proof of concept code repeated all queries for which a spike
was detected until a single query was left. This allowed to reveal an
ephemeral port with a high success rate.

\begin{figure}[h!]
\begin{center}
  \includegraphics[width=0.99\columnwidth,clip]{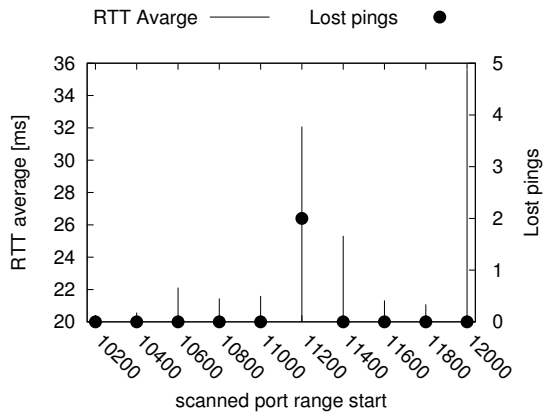}
\end{center}
\caption{The range scan of an idle connection. Spoofed segments are
  covering 200 port ranges. 5 pings and 6000 spoofed segments (30 to
  each port) are sent for each range. RTT and loss rate spikes reveal
  the ephemeral port is somewhere between 11200 and 11400.}
\label{port_scan_range}
\end{figure}

The lower bound on a query time is a single ping RTT, because at least
one ping needs to be sent to determine the query result. Even for a
relatively short RTT of 20ms, if a full range of 64k ephemeral ports
needs to be scanned, the sequential scan would require at least 21
minutes. When bandwidth from the attacker to the victim is large,
continuous range of ports can be probed in each sequence of spoofed
segments. Such sequence can be interpreted as a query 'Is the
connection using a port between X and Y?'. If a part of the sequence
is reflected, the answer is yes, and a sequential search can be used
to find the exact port number. In the experimental setup such range
queries worked well and considerably reduced time of the scan (see
figure~\ref{port_scan_range}). Table~\ref{port_results} summarizes
experimental results. The results were similar for both tested
systems. The attacker can further improve performance if the targeted
connection uses ephemeral port from a smaller range.

\begin{table*}
\caption{Ephemeral port search. The full space of 65k ephemeral ports
  was searched.}
\label{port_results}
\begin{center}
\begin{tabular}{l|l|l|l|l|l|l}
connection&scan   &queries&pings        &max ports &spoofed        &reflected\\
type      &time[s]&       &             &per query &segments       &segments\\
\hline
\hline
idle      & 35    &592    &  2960       &   200    &2202780        &  330\\
          &       &       & 5/query     &          &30/port/query  &25kB\\
          &       &       & 0.25MB total&          &171MB total    &\\
          &       &       & 22ms avg RTT&          &               &\\
\hline
download  &852    &849    &  8490       &  100     &73614000       &  12000\\
          &       &       & 10/query    &          &1000/port/query&0.9MB\\
          &       &       & 0.7MB total &          & 5741MB total  &\\
          &       &       &749ms avg RTT&          &               &\\
\hline
upload    &690    &852    & 8520        &  100     &74013000       &  10000\\
          &       &       & 10/query    &          &1000/port/query&0.8MB\\
          &       &       & 0.7MB total &          & 5773MB total  &\\
          &       &       &656ms avg RTT&          &               &\\
\hline
\end{tabular}
\end{center}
\end{table*}

\subsubsection{A side note on Netfilter}

It is interesting why Netfilter does not drop SYN-ACK segments arriving
in a context of an already established connection. There are at least
two signals that indicate a SYN-ACK segment is incorrect: 1. ACK
number does not acknowledge any SYN segment, 2. Data was already
exchanged in both directions, three way handshake must have had
finished successfully. A comment in the Netfilter source code says
\textit{'Our connection entry may be out of sync, so ignore packets
  which may signal the real connection between the client and the
  server'} (ignore here means do not drop). The problem is that
Netfilter is a completely separate layer from the Linux TCP stack. It
does not have access to the real state of a TCP connection, but
recreates it based on segments it has seen. It does not assume the
protected end point is on the same machine and that segments it has
accepted reached the destination. For these reasons, tracking state of
a TCP connection and determining if a segment can be safely dropped is
very complex. As demonstrated in~\cite{ipfilter}, there are many
corner cases to consider that can lead to hanged connections when
handled incorrectly.

\subsection{Sequence numbers}

To inject data at the start of a window of a one end of the connection
(the victim or its peer), the attacker needs to know the sequence
number of the next octet to be sent (SND.NXT) by the other end. The
exact value of the SND.NXT of the end point to which data is inserted
does not need to be known, it is enough that the segment that injects
data has an acceptable acknowledge number set. Injecting data is
relatively easy if the end point is not actively receiving data from
its peer. If it is not the case, the window and SND.NXT constantly
change, introducing an additional obstacle that the attacker needs to
overcome. The paper does not try to address these difficulties.

Steps needed to determine SND.NXTs significantly differ for the two
tested systems. In both cases ACK segments with an ephemeral port
determined in the previous step are used. Windows firewall never drops
ACK segments that are exchanged withing an established connection
(have correct IP addresses and ports), so only rules defined in RFC
793 need to be taken into account when analyzing Windows
responses. Netfilter implements stricter filtering rules. The following
subsections demonstrate that stricter filtering can significantly
reduce resources needed by the attack.

\subsubsection{Host strictly following RFC 793}

The sequence number of the victim's peer needs to be determined
first. If a sequence number of an incoming ACK segment is in window,
and an acknowledge number is acceptable, the segment does not trigger
any response. Otherwise, ACK segment is sent in response. According to
RFC 793, acknowledge number is acceptable if it is equal to the next
octet to be sent or lower by at most $2^{31}$. In other words, an
acceptable acknowledge number lies in range: [$\text{SND.NXT} -
  2^{31}$, SND.NXT] (using the 'sequence space arithmetic').  Because of
this, out of two acknowledge numbers that differ by $2^{31}$ one is
guaranteed to be acceptable. The attacker needs to send
\[N=\frac{2\times2^{32}}{\text{window size}}\] queries to find in-window
sequence number. The risk of accidentally corrupting the session is
negligible. The session would be corrupted only if the attacker
happens to acknowledge data that was lost in transit. In such case the
data won't be retransmitted.

The attacker does not need to know the size of the victim's window,
although it can be often easily determined
(see~\cite{fingerprinting}). The attacker can first assume the maximum
allowed window (1GB) and try sequence numbers that differ by
$2^{30}$. If none of such sequence numbers is in-window, the
attacker can try sequence numbers in the middle of ranges probed in
the previous step. If the victim uses 0.5GB window, one of such
sequence numbers should be in-window. The steps can be repeated,
each time the assumed window size is divided by two until in-window
sequence number is found. Such search is described with more details
in~\cite{ip_id_tcp_attack}.

Out of $N$ queries, a single one that does not generate a positive
response needs to be found. The situation is opposite to the port
scanning, where a single query that does generate a response was
searched for. In practice, searching for a negative answer is more
difficult:

\begin{itemize}
  \item Bottleneck is constantly overloaded. Scanning needs to be done
    in sequence, with long enough intervals between subsequent queries
    for a bottleneck's queue to empty. Scanning several values at once
    is not possible - it is relatively easy to distinguish between
    a traffic spike and a lack of traffic spike, it is much harder to
    distinguish between a traffic spike and a slightly smaller traffic
    spike.
  \item Natural traffic may mask the lack of response. In contrast,
    when query to which the system responds is searched for, natural
    traffic can only magnify the traffic spike.
\end{itemize}

\begin{figure}[h!]
\begin{center}
\subfigure[
  Connection idle, 5 pings/seq number, 30 spoofed segments/seq number]{
  \includegraphics[width=0.99\columnwidth,clip]{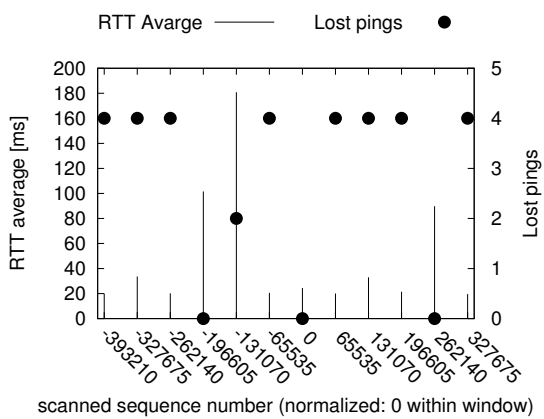}
}
\subfigure[
  Connection uploading data,
  10 pings/seq number, 1000 spoofed segments/seq number]{
  \includegraphics[width=0.99\columnwidth,clip]{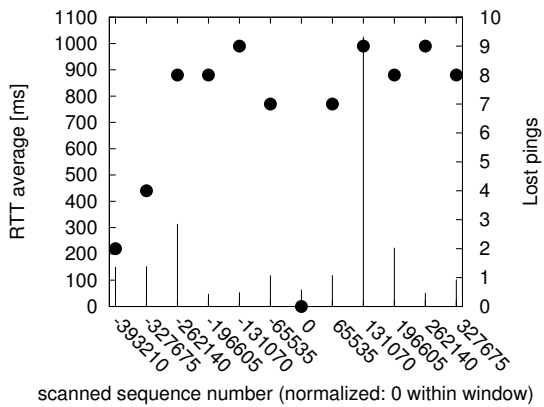}
}
\end{center}
\caption{When a sequence number of a spoofed segment is in-window,
  the smallest average RTT and loss rate are measured. The RTT and
  loss rate of other probes are large, increasing duration of
  the scan. Natural traffic spike could mask a minimum. Ping is
  considered lost if the response did not arrive within two RTTs of
  previous pings related to the same sequence number.}
\label{sqn_scan}
\end{figure}

Figure~\ref{sqn_scan} illustrates how RTT decreases when a probed
sequence number is within window. Table~\ref{sqn_results} shows that
even in case of an idle connection, the time needed for a scan to
finish is significant. In the experimental setup, the PoC code would
need roughly about 36 hours to complete a sequential scan of a
connection uploading data.

\begin{table*}
\caption{In-window sequence number search for the host strictly
  following RFC 793 processing rules. The host used a window of
  65kB.}
\label{sqn_results}
\begin{center}
\begin{tabular}{l|l|l|l|l|l}
connection&scan    &queries&pings        &spoofed    &reflected \\
type      &time[s] &       &             &segments   &segments  \\
\hline
\hline
idle      & 16860  &131338 & 394014      &3940140    &3939930   \\
          &$\sim$5h&       & 3/query     &30/query   &\\
          &        &       & 32MB total  &307MB total&          \\
          &        &       & 85ms avg RTT&           &          \\
\hline
\end{tabular}
\end{center}
\end{table*}

Knowing in-window sequence number, the attacker can find the victim's
peer SND.NXT by looking for the lowest sequence number that does not
generate a response. Such value is at most window size before the
in-window sequence number and can be found with binary search in
$log(\text{window size})$ queries.

Also, if the value of the victim's SND.NXT is needed, it can be now
easily determined. 31 queries are required to binary search for the
highest acknowledge number that does not generate any response (is
acceptable). This number is equal to the SND.NXT of the victim.

\subsubsection{Host protected by Netfilter}

The sequence number of the victim is determined first. The technique
exploits acknowledge number validation rules described in
section~\ref{ephemeral_port}. An ACK segment is accepted only if its
acknowledge number is equal to the next octet to be sent or lower by
at most $\max(66000, \text{largest sender window seen})$. In other
words, an acceptable acknowledge number lies in range:
[$\text{SND.NXT} - \max(66000, \text{largest sender window seen})$,
  SND.NXT] (using the 'sequence space arithmetic'). A segment with not
acceptable acknowledge number is silently dropped by the firewall.
Netfilter does not validate sequence numbers of ACKs. Linux TCP layer
to which not dropped segments are passed, validates a sequence number
and responds with ACK if it is out of a window.

This allows to find an acceptable acknowledge number in
$2^{32}/\max(66000, \text{largest sender window seen})$ tries. If the
victim responds to a segment that had an incorrect sequence number, it
means the acknowledge number was accepted by Netfilter. Searching
for an acceptable acknowledge number is analogous to searching for an
ephemeral port. At most $2^{32}/66000 = 65075$ values need to be
probed and only one of these values generates a positive response,
which allows to probe several values in a single query. See
table~\ref{port_results} again for estimates of resources needed.

There is a trick that allows to further improve the search efficiency.
Netfilter can be easily fooled to set the value of the 'largest sender
window seen' to the maximum value allowed by a window scaling factor
that was set during the connection establishment. To do it, 65075 ACKs
need to be sent, covering the whole $2^{32}$ acknowledge number space
with values that differ by 66000. All these ACKs need to have window
size set to the maximum allowed value: $0xFFFF$. One of the ACKs
should be accepted by  Netfilter and sets maximum window seen so
far to $0xFFFF \times 2^{\text{window scaling factor}}$ (note that
this does not affect the real window size, the TCP end point rejects the
ACK because it carries an incorrect sequence number). In the
experimental setup, the sender set the window size to 114 with the
scaling factor of 7, which resulted in a small window of
$114\times2^7=14592$B. The sequence of spoofed ACKs fooled
Netfilter that the window increased to
$0xFFFF\times2^7=8388480$B. Such a window allowed to cover the whole
$2^{32}$ acknowledge number space with only 512 values. As it was the
case when the host following RFC 793 was targeted, the attacker
does not need to know the size of the victim's window and the scaling
factor. Maximum allowed window of 1GB can be assumed, and divided by
two until an acceptable acknowledge number is found.

See table~\ref{ack_results} for summary of resources needed by the search.

\begin{table*}
\caption{Acceptable acknowledge number search. The attacker fooled
  Netfilter that the sender window increased to 8.3MB, this allowed to
  cover the whole acknowledge number space with a small number of queries.}
\label{ack_results}
\begin{center}
\begin{tabular}{l|l|l|l|l|l|l}
connection&scan   &queries&pings        &max ack values&spoofed    &reflected\\
type      &time[s]&       &             &per query &segments       &segments\\
\hline
\hline
idle      & 3.4   &60     & 300         &   25     &20520          &240\\
          &       &       & 5/query     &          &30/ack value/query &18kB\\
          &       &       & 24kB total  &          &1.6MB total    &\\
          &       &       & 21ms avg RTT&          &               &\\
\hline
download  &61     &51     & 510         &  25      &627000          &4000\\
          &       &       & 10/query    &          &1000/ack value/query&0.3MB\\
          &       &       & 42kB total  &          &49MB total      &\\
          &       &       &866ms avg RTT&          &                &\\
\hline
upload    &59     &56     & 560         &  25      &728000          &6000\\
          &       &       & 10/query    &          &1000/ack value/query&0.5MB\\
          &       &       & 45kB total  &          &57MB total      &\\
          &       &       &602ms avg RTT&          &                &\\
\hline
\end{tabular}
\end{center}
\end{table*}

Knowing an acceptable acknowledge number, binary search can be used to
find the sequence number of the next octet to be sent by the
victim. This requires $\log(\max(66000, \text{largest sender window
  seen}))$ queries.

If the victim's peer SND.NXT needs to be known, the attacker
has several ways to reveal it:
\begin{itemize}
  \item Segments with a single byte of data can be used. Netfilter
    validates sequence numbers of segments that carry data. If the
    number is in-window, the segment is passed to the TCP layer
    which generates ACK in response because data is out of order. If
    the sequence number is out of the window, Netfilter drops the
    segment. This technique carries the risk of corrupting the session
    with the accepted byte.
  \item If the attacker is able to send spoofed traffic to both ends,
    and to reliably monitor traffic spikes of both ends, the other end
    of the connection can be targeted. If the other end follows the
    RFC 793, only 32 queries are needed to find the second SND.NXT. If
    it is protected by Netfilter, the steps described in this section
    can be used again.
  \item Resource intensive search for a sequence number that does not
    generate any response can be performed in a similar way it was
    done for a system following RFC 793 in the previous
    subsection. The only difference is that acceptable acknowledge
    number is already known, only in-window sequence number needs to
    be found.
\end{itemize}

\subsection{Other variants}

Different TCP stacks may implement different segment processing rules,
possibly closing some leaks described in this paper, or opening new
ones. For example, to prevent a blind RST injection attack described
in~\cite{slipping}, a new recommendation for RST processing was
created~\cite{blind_robust}. According to RFC 793 any in-window RST
should be accepted and should reset the connection. The stricter and
safer rules require RST to have sequence number exactly equal to the
next expected sequence number, otherwise, in-window RST segment should
generate ACK in response without resetting the connection. The
document advises to optionally throttle such ACKs. If such ACKs are
not throttled, the attacker can query for a window using RST segments
with little risk of accidentally resetting the connection.

\section{Advanced scanning technique}

\label{advanced_technique}
TCP Fast Retransmit~\cite{fast_rfc} can be exploited to trigger
substantial traffic spikes with relatively small number of spoofed
segments. Fast Retransmit is activated by 3 duplicated ACKs. TCP layer
interprets such duplicates as a message that a segment was lost but 3
subsequent segments successfully arrived at the destination. Each
following duplicated ACK is interpreted as an acknowledgement that
another segment was successfully received, but the lost segment still
didn't reach the destination. Because segments are successfully
leaving the network, sender sends a new segment in response to each
such duplicated ACK. A burst of ACKs can trick the sender to send a
full window of data in a very short time as described
in~\cite{congestion}. The amplification factor for a network with MTU
1500 is 37. This allows the attacker to trigger observable traffic
spikes with much fewer spoofed segments.

The technique can also be used to detect ephemeral port number of a
host that does not filter segments addressed to not existing
connection but responds with RST to each such segment. A sequence of
spoofed segments directed to an incorrect port, results in a sequence
of RSTs that are silently dropped by the other end point with no side
effect. A sequence of spoofed segments directed to the correct port,
results in a sequence of ACKs that trigger the other side to abruptly
send a full window of data. If the attacker can detect the spike in
traffic caused by this window of data, the port can be determined.

The technique was not tested in practice.

\section{Protection}

To be fully protected against side channel information leakage
described in this paper, the protocol would need to ensure that not
authenticated segments are never answered. If it was the case, the
only information that would leak to an off-path attacker, would be
that the segment was not authenticated. Providing authentication
mechanism is strong enough to make probability of generating
acceptable request negligible, the attacker learns nothing through the
side channel that couldn't be figured out without mounting the attack.
The TCP Authentication Option~\cite{auth_rfc} provides exactly such
mechanism, but the option is not widely used.

In case of sequence numbers based authentication, it can be difficult to
ensure in a backward compatible way that the protocol never responds
to rejected segments. Sequence numbers have double purpose. They were
intended primary for detecting duplicates, lost and out of order
segments. The use of sequence numbers as a protection mechanism
against an adversary was emergent, not even mentioned in the original
specification. If Netfilter filtered SYN-ACK segments addressed to an
established connection and dropped ACK segments with invalid sequence
numbers, the attack against a system protected by Netfilter would be
probably impossible. But such stricter filtering rules require very
careful analysis to prevent hanged connections in corner cases.

Throttling responses to rejected segments should be sufficient to
make the information leakage non-exploitable in practice.
Throttling mechanism for ACKs generated in response to in-window
RSTs and in-window SYN-ACKs was proposed in~\cite{blind_robust}. To be
effective, the mechanism would need to throttle also ACKs generated in
response to other rejected segments.

The attack is the easiest if the attacker shares an edge router
with the victim. The first few hops are also the best place to
reliably filter spoofed IP packets. Network that is configured to drop
such traffic is protected at least against a local attacker.

Queueing policy that better isolates traffic coming from different
users could make the attack more difficult to execute. A privacy
protecting scheduling policy was studied
in~\cite{privacy_routing}. The authors were able to significantly
reduce the correlation between traffic patterns of users sharing a
routing queue without introducing prohibitive performance
degradation. The designed policy reduced the leakage of information
regarding the traffic pattern of a user, but the traffic load of a
user was still leaking through the increased packet processing
time. To execute the attack described in this paper, it is enough to
detect increased traffic load, knowing the exact traffic pattern is
not necessary. Further research is needed to assess the effect of
different queuing policies on the attack applicability.

\section{Summary}

The paper demonstrated how changes in processing time of packets that
traverse a shared queue can reveal if a host responded to spoofed
traffic. It was shown that in case of the TCP protocol, being able
to determine if a system responded to spoofed segments is sufficient
to compromise the session, direct interception of the TCP traffic is not
required. Two different TCP implementations with different processing
rules were examined. Both implementations responded to partially
incorrect TCP segments, allowing the attacker to determine values of
secret fields in separate steps. Substantial part of the work was
dedicated to experiments to determine if the attack is practical in
real-life scenario and to provide estimates of resources needed. The
paper concluded with the discussion of possible attack prevention
mechanisms.

The work did not try to determine the practical limits of the
technique. There is a lot of room for further experiments in scenarios
more adverse for the attacker (lower bandwidth between the attacker
and the victim, busy bottleneck shared between many users, different
queuing policies). Provided proof of concept can be used as a starting
point for such experiments. The paper also did not attempt to provide
a detailed survey of applicability of the technique against popular
TCP implementations. Finally, the paper concentrated on compromising
TCP session, but the presented technique can be applicable in other
scenarios.

\section{Acknowledgments}

The author would like to pass a non-spoofed ACK to \mbox{Wojtek}
\mbox{Matyjewicz} for reviewing first drafts of the paper, valuable
comments and discussions.

\end{document}